\documentclass[12pt, prd, showpacs]{revtex4}
\usepackage{amssymb}
\usepackage{amsmath}

\input{tcilatex}

\begin{document}

\title{Ultrahigh energy particle collisions near the black hole horizon in
the strong magnetic field}
\author{O. B. Zaslavskii}
\affiliation{Department of Physics and Technology, Kharkov V.N. Karazin National
University, 4 Svoboda Square, Kharkov 61022, Ukraine}
\email{zaslav@ukr.net }

\begin{abstract}
We consider collision between two charged (or charged and neutral) particles
near the black hole horizon in the strong magnetic field $B$. It is shown
that there exists a strip near the horizon within which collision of any two
such particles leads to ultrahigh energy in the centre of mass frame. The
results apply to generic (not necessarily vacuum) black holes.
\end{abstract}

\maketitle

\section{Introduction}

If two particles move towards a black \ hole and collide near the horizon,
under certain additional conditions their energy $E_{c.m.}$ in the centre of
mass can become unbound. There are different scenarios of this kind: a black
hole should be rotating \cite{ban}, electrically charged \cite{jl} or
immersed in the magnetic field \cite{fr}. In the latter case, in the
situation considered previously, collision occurs near the innermost stable
circular orbit (ISCO) \cite{fr} that lies near the horizon for the magnetic
field strength $B$ large enough \cite{ag}. Formally, $E_{c.m.}\rightarrow
\infty $ requires $B\rightarrow \infty $ in this scenario. In doing so, the
individual angular momentum on ISCO also grows unbound.

To realize this scenario, it is sufficient to take the simplest case of the
Schwarzschild black hole, so the magnetic field affects motion of particles
but not the metric itself \cite{fr}. The corresponding approach was
generalized to the case of the Kerr metric \cite{weak}.

The aim of the present letter is to draw attention to one more mechanism.
Similarly to \cite{fr} and \cite{weak}, it requires the strong magnetic
field. However, its realization is not connected with ISCO. In the very
vicinity of the horizon it becomes a universal phenomenon and works for
particles with arbitrary individual energies and angular momenta.

Throughout the paper we use units in which fundamental constants are $G=c=1$.

\section{Basic equations}

To simplify matter, let us consider the static spherically symmetric metric
of the form

\begin{equation}
ds^{2}=-N^{2}dt^{2}+\frac{dr^{2}}{N^{2}}+r^{2}(d\theta ^{2}+\sin ^{2}\theta
d\phi ^{2})\text{,}  \label{m}
\end{equation}%
where $N$ depends on $r$ only. The horizon lies at $N=0$. If $N^{2}=1-\frac{%
2M}{r}$ we return to the case of the Schwarzschild black hole \cite{fr}. In
general, we do not specify the form of $N$ explicitly and even do not
require it to be unaffected by the magnetic field. In particular, we do not
require $Br_{+}\ll 1$ ($r_{+}$ is the horiozn radius, $B$ is the effective
magnetic field), allowing $Br_{+}\sim 1$. In eq. (\ref{m}), the metric
coefficients satisfy the relation $g_{rr}g_{00}=-1$ but this is a weak
restriction that somewhat simplifies formulas without the loss of generality
for the effect under discussion.

We assume that there is a vector-potential that has the only nonvanishing
component%
\begin{equation}
A^{\phi }=\frac{B}{2}\text{,}  \label{a}
\end{equation}%
where $B$ is, in general, the function of $r$. In the vacuum case, the
Maxwell equations are satisfied with $B=const$ \cite{wald}. However, we
consider a more general case of "dirty" (surrounded by matter) black holes.

Let a particle with the mass $m$ and electric charge $q$, move in the
background (\ref{m}) with the vector-potential (\ref{a}). The kinematic
momentum $p_{\mu }=mu_{\mu }$ and the generalized one $P_{\mu }$ are related
according to $p_{\mu }=P_{\mu }-qA_{\mu }$. Here, $u^{\mu }=\frac{dx^{\mu }}{%
d\tau }$ is the four-velocity, $\tau $ is the proper time. The components $%
P_{t}=-E$ and $P_{\phi }=L$ are conserved, $E$ having the meaning of the
energy, $L\,$\ being the angular momentum. Then, equations of motion read%
\begin{equation}
m\dot{t}=\frac{E}{N^{2}}
\end{equation}%
\begin{equation}
m\dot{\phi}=\frac{L}{r^{2}}-q\frac{B}{2}\text{,}
\end{equation}%
\begin{equation}
m\dot{r}=\varepsilon Z\text{,}
\end{equation}%
where%
\begin{equation}
Z=\sqrt{E^{2}-m^{2}N^{2}(1+\beta ^{2})}\text{,}  \label{z}
\end{equation}%
\begin{equation}
\beta =\frac{L}{mr}-b\text{, }b=\frac{qBr}{2m}\text{.}
\end{equation}%
We assume that a black hole is electrically neutral.

Let two particle 1 and 2 with masses $m_{1}$ and $m_{2}$, electric charges $%
q_{1}$ and $q_{2}$ and four-velocities $u_{1}^{\mu }$ and $u_{2}^{\mu }$
collide in some point. One can define in the same point the energy $E_{c.m.}$%
in the centre of mass frame (CM frame) according to%
\begin{equation}
E_{c.m.}^{2}=-(m_{1}u_{1}^{\mu }+m_{2}u_{2}^{\mu })(m_{1}u_{1\mu
}+m_{2}u_{2\mu })=m_{1}^{2}+m_{2}^{2}+2m_{1}m_{2}\gamma \text{,}
\end{equation}%
where%
\begin{equation}
\gamma =-u_{1}^{\mu }u_{2\mu }
\end{equation}%
is the Lorentz factor of relative motion.

Using equations of motion, one can find%
\begin{equation}
m_{1}m_{2}\gamma =\frac{E_{1}E_{2}-\varepsilon _{1}\varepsilon _{2}Z_{1}Z_{2}%
}{N^{2}}-m_{1}m_{2}\beta _{1}\beta _{2}\text{.}  \label{ga}
\end{equation}

As one approaches the horizon, $N\rightarrow 0$. Then, for head-on collision
($\varepsilon _{1}\varepsilon _{2}=-1$) the Lorentz factor $\gamma
\rightarrow \infty $ for any values of the angular momentum and magnetic
field. The rotational analogue of this phenomenon was studied in \cite{pir1}
- \cite{pir3}. Hereafter, we consider the case $\varepsilon _{1}=\varepsilon
_{2}=-1$, so both particles move towards the horizon. In general, $\gamma $
remains finite near the horizon, and the question is whether and how $\gamma 
$ can become unbound.

\section{Ultrahigh energy collisions}

For any finite values of all relevant quantities, one can calculate the
horizon limit $N\rightarrow 0$ of (\ref{ga}) and find that it is finite.
More precisely,%
\begin{equation}
\gamma _{0}\equiv \lim_{N\rightarrow 0}\gamma =\frac{1}{2}[\frac{%
m_{1}E_{2}[1+\beta _{1H}^{2})}{m_{2}E_{1}}+\frac{m_{2}E_{1}(1+\beta
_{2H}^{2})}{m_{1}E_{2}}]-\beta _{1H}\beta _{2H}=\frac{1}{2}\frac{(\alpha
_{1}\beta _{2H}-\alpha _{2}\beta _{1H})^{2}}{\alpha _{1}\alpha _{2}}+\frac{%
\alpha _{1}}{2\alpha _{2}}+\frac{\alpha _{2}}{2\alpha _{1}}\text{,}
\label{lim}
\end{equation}%
where subscript "H" means that the corresponding quantity is calculated on
the horizon,%
\begin{equation}
\alpha _{1}=\frac{E_{1}}{m_{1}}\text{, }\alpha _{2}=\frac{E_{2}}{m_{2}}\,%
\text{.}
\end{equation}

However, the situation can change if (i) collisions occur not exactly on the
horizon but in its vicinity at $r=r_{c}\approx r_{H}$, so $N(r_{c})\equiv
N_{c}\ll 1$, (ii) the quantity $\beta _{iH}\gg 1$, (iii) the factors (i) and
(ii) are related to each other in such a way that%
\begin{equation}
N_{c}\beta _{iH}\equiv s_{i}\sim 1\text{, }  \label{s}
\end{equation}%
where $i=1,2.$

Then, the Lorentz factor 
\begin{equation}
\gamma \approx \frac{F}{N_{c}^{2}}\text{, }F=\alpha _{1}\alpha _{2}-\sqrt{%
\alpha _{1}^{2}-s_{1}^{2}}\sqrt{\alpha _{2}^{2}-s_{2}^{2}}-s_{1}s_{2}.
\label{gas}
\end{equation}

The numerator is positive, except from for the particular case $\alpha
_{1}s_{2}=\alpha _{2}s_{1}$ when $F$ vanishes. We exclude this case from
consideration. Then, $\gamma $ can become as large as one likes. For $%
s_{i}\ll \alpha _{i}$, eq. (\ref{gas}) turns into (\ref{lim}) in the main
approximation, if finite corrections in (\ref{lim}) are neglected.

As we must have $Z_{i}\geq 0$, this mechanism works in the immediate
vicinity of the horizon only, so 
\begin{equation}
0<N_{c}\leq \frac{E_{i}}{m_{i}\beta _{iH}}  \label{str}
\end{equation}%
or, equivalently, $\alpha _{i}\geq s_{i}$. It is curious that, formally, the
restriction (\ref{str}) on the size of the strip within which $E_{c.m.}$ is
ultra-high, resembles the corresponding restriction (18) in \cite{gp} or
(18) in \cite{prd}, for rotating nonextremal black holes without the
magnetic field.

Using the algebraic inequality%
\begin{equation}
\frac{1}{2}\frac{(\alpha _{1}s_{2}-\alpha _{2}s_{1})^{2}}{\alpha _{1}\alpha
_{2}}\leq \alpha _{1}\alpha _{2}-\sqrt{\alpha _{1}^{2}-s_{1}^{2}}\sqrt{%
\alpha _{2}^{2}-s_{2}^{2}}-s_{1}s_{2}\text{,}
\end{equation}%
one can easily show that $\gamma _{0}<\gamma (N_{c})$ with $N_{c}>0$. In
other words, to gain the maximum possible $E_{c.m.}$, it is more profitable
to arrange collision not on the horizon itself but in its vicinity,
notwithstanding the fact that high $E_{c.m.}$ arise just due to the horizon!
This circumstance is similar to the observation made in \cite{piat} for
rotating nonextremal Kerr black holes and extended in \cite{circ} for
generic dirty rotating black holes.

In the particular \ case when particle 2 is neutral, $s_{2}=0$. Then, it is
seen from (\ref{gas}) that for a given $\alpha _{1}$, $\alpha _{2}$, the
function $F$ attains it maximum value if $s_{1}=\alpha _{1}$ that
corresponds to the turning point (cf. \cite{piat}, \cite{circ}).

There are two ways to achieve large value of $\beta _{1}$. The first one is
to increase $L_{1}\,.$ (High energy collisions with indefinitely large $L_{1}
$ near the horizon were considered in \cite{nn1} (case 3) but with
additional assumptions that the radial velocity is vanishingly small.)
However, this imposes rather severe restriction on $L_{1}$ that represents
some problem for the realization of such a scenario. Meanwhile, there is
more physical way to achieve large $\beta _{iH}\,\ $since it is possible to
take a large magnetic field, so that%
\begin{equation}
b_{iH}\gg 1\text{.}  \label{b}
\end{equation}%
Then, $\beta _{iH}\sim B$, and it follows from (\ref{gas}) that%
\begin{equation}
\gamma \sim B^{2}\text{.}
\end{equation}

For the effect to occur, at least one of particle should be electrically
charged. If they both are neutral, there is no interaction with a magnetic
field, and $B$ does not enter the expression for $\gamma $ (\ref{z}), (\ref%
{ga}) at all.

It is important that if (\ref{b}) is obeyed, the relations (\ref{s}), (\ref%
{gas}) are satisfied for particles with \textit{any} finite values of the
energy and momentum. Therefore, the phenomenon under discussion acquires
universal character. It turns out that in the vicinity of the horizon in a
strong magnetic field, collisions between any two particles with arbitrary
energies and angular momenta give rise to ultra-high value of $E_{c.m.}$ !

\section{Summary}

Thus we considered particle collisions in the strong magnetic field near the
black hole horizon. It turned out that in the immediate vicinity of the
horizon, $E_{c.m.}$ grows as $B^{2}$. For comparison, in the Frolov's
process \cite{fr}, $E_{c.m.}\sim B^{1/4}$ (see also eqs. 62, 63 of \cite%
{weak}), so the present mechanism is more efficient. In both cases there is
restrictions on the location of collision. In our scenario, it should happen
near the horizon within the coordinate distance determined by (\ref{str}).
In the Frolov's case collision occurs near ISCO. It is also worth noting
that $L_{1}\sim B$ for the particle on ISCO (see eq. 49 of \cite{weak}) but $%
L_{1}$ and $L_{2}$ are arbitrary finite quantities in our scenario.

The results apply to any dirty static black hole since the condition of
spherical symmetry can be relaxed easily without the loss of generality.
Also, it admits straightforward extension to rotating black \ holes. In
doing so, there is no need to invoke additional assumptions that the
background metric is almost unaffected by the magnetic field since the basic
formulas like (\ref{gas}), (\ref{str}) work anyway.

Now, on the basis of \cite{fr}, \cite{weak} and the present work, we can
conclude that there exist at least two main scenarios of ultrahigh energy
collisions in the magnetic field near black holes. They include collision
(i) in the strong field on ISCO, (ii) in the strong field in the immediate
vicinity of the horizon. In the present work, we described scenario (ii).
The distinctive feature of this scenario consists in that it does not
include dynamic characteristics of particles, so their energy and angular
momentum can be arbitrary finite quantities.

Thus once there is a very strong magnetic field near the black hole horizon,
one can always find the strip around the horizon where this phenomenon does
occur. This lends the property of universality to the scenario under
discussion and enlarges hopes to find realization of ultra-high energy
collisions in nature.

\end{document}